\documentclass[12pt,a4paper]{article}
\usepackage{amsfonts,latexsym}
\usepackage{amsmath,amssymb}
\usepackage{graphicx,color}

\renewcommand{\thefootnote}{\fnsymbol{footnote}}

\begin{document}

\vspace{12mm}

\begin{center}
{{{\Large {\bf Photon ring bounds of scalar hairy charged black holes }}}}\\[10mm]

{Yun Soo Myung\footnote{e-mail address: ysmyung@inje.ac.kr}}\\[8mm]

{Institute of Basic Sciences and Department  of Computer Simulation, Inje University, Gimhae 50834, Korea\\[0pt] }

\end{center}
\vspace{2mm}

\begin{abstract}
We study  photon rings of constant scalar hairy black holes with mass $m$, charge $Q$, and scalar hair $S$  obtained from the Einstein-Maxwell-conformally coupled scalar  theory.
These black holes are classified as  scalar hairy  Reissner-Nordstr\"{o}m (SHRN)  black hole, scalar hairy charged black hole with  $S> -Q^2$, and mutated-RN black hole.
The first two respect both dominant and strong energy conditions and have positive ADM mass and entropy, while the last does not respect two energy conditions and has negative ADM mass and entropy.
We find that all of these black holes respect the lower bound ($r_\gamma\ge 1.2 r_+$) of photon rings.
We obtain  all real bounds of photon rings  for   these black holes and discuss physical and observational properties of real bounds.
It is shown that the observationally favored  region  based on the shadow radius includes  SHRN black hole, scalar hairy charged black hole, and mutated-RN black hole.
\end{abstract}
\vspace{5mm}

\vspace{1.5cm}

\hspace{11.5cm}{Typeset Using \LaTeX}
\newpage
\renewcommand{\thefootnote}{\arabic{footnote}}
\setcounter{footnote}{0}

%%%% Introduction %%%%

\section{Introduction}

No-hair theorem states  that an asymptotically flat black hole is completely described by mass $M$, electric charge $Q$, and angular momentum $J$~\cite{Ruffini:1971bza}.
In this direction, a minimally coupled scalar does not obey the Gauss-law and thus, a black hole is unlikely to have any scalar hairs~\cite{Herdeiro:2015waa}. On the other hand, introducing the Einstein-conformally coupled scalar  theory, one has found the Bocharova-Bronnikov-Melnikov-Bekenstein (BBMB) black hole with mass $m$ and scalar hair $\phi(r)$ which blows up on the horizon~\cite{Bocharova:1970skc,Bekenstein:1974sf}. This is the first counterexample to the no-hair theorem.
However,  thermodynamical properties of the  BBMB black hole  are not promising  because the Hawking temperature is always zero and its entropy becomes infinite~\cite{Winstanley:2004ay,Karakasis:2021rpn}. Also, it turned out that the BBMB black hole is unstable against perturbations~\cite{Kobayashi:2014wsa}. Its series (numerical) solution was found in~\cite{Myung:2019adj}.  For the BBMB black hole, the photon ring (null unstable  circular geodesics) is fixed  as $r_\gamma=2m$ where the effective Newton  constant diverges~\cite{Shinohara:2021xry}. The uniqueness for the outside region  of the photon ring  has been proved ~\cite{Tomikawa:2017vun,Yoshino:2017gqv}.

It is well known that closed photon rings  are of importance in determining  the physical and observational properties of black hole spacetimes~\cite{Stefanov:2010xz,EventHorizonTelescope:2019dse,Cunha:2020azh}.
Very recently, it was shown that the lower bound on the photon ring radius $r_\gamma$  of black holes is determined  by $r_\gamma\ge 1.2r_+$ with $r_+$ horizon radius if one chooses  Einstein gravity  with a traceless energy-momentum tensor~\cite{Hod:2023jmx}.
Two examples to be tested are recommended as Reissner-Nordstr\"{o}m (RN) and Einstein-Yang-Mills black holes.
In the case of RN black hole, its real bound is given by $1.5\le r_\gamma/r_+ \le 2$. Therefore, the RN black hole respects the lower bound ($r_\gamma\ge 1.2r_+$).
However, it is not easy to test  Einstein-Yang-Mills black holes because their solutions were known numerically~\cite{Bizon:1990sr,Winstanley:1998sn,Bjoraker:2000qd,VanderBij:2001ia,Moon:2011hq}.

Photon rings of constant scalar hairy charged black holes~\cite{Astorino:2013sfa} were extensively  used to describe the shadow of M87* supermassive black hole~\cite{Khodadi:2020jij} and gravitational lensing effects of M87* and SgrA* supermassive black holes~\cite{QiQi:2023nex}.
It is important to  note that these black holes are found from the Einstein-Maxwell-conformally coupled scalar (EMCS) theory whose stress-energy tensor is traceless on-shell configuration.
So, these constant scalar hairy black holes could  be another  candidate for testing the photon ring  bounds. We note that the shadow of non-charged black hole with constant scalar hair was  employed to constrain the scalar hair $S$ when comparing with the resent EHT results~\cite{Vagnozzi:2022moj}.

In this work, we wish to employ   constant scalar hairy charged black holes  with mass $m$, charge $Q$, and scalar hair (charge) $S$ to derive  the real bounds for the ratio of  photon ring radius  to horizon radius.
The scalar hair `$S$' plays an important role in matching the observational data. 
According to photon ring bounds, these black holes with mass $m=1$ are classified into  scalar hairy  RN (SHRN) black holes with $0<S<1$ and $0<Q<\sqrt{1-S}$,  scalar hairy charged black hole with  $S>-Q^2$,
and mutated-RN black hole (Einstein-Rosen bridge) with $S<-Q^2$~\cite{Chowdhury:2018pre}.
We note here that the case of  $Q=\sqrt{-S}$ denotes Schwarzschild-like black hole differing from Schwarzschild black hole with $S=Q=0$ and the case of $Q=\sqrt{1-S}$ represents  extremal black hole with $m=1$ differing from extremal RN black hole with $S=0$. We obtain  all real bounds of photon rings  for   these black holes, and discuss physical and observational properties of real bounds by comparing the EHT (M87*) results.

\section{EMCS theory and its black hole solution}
We start with the EMCS  theory given by
\begin{eqnarray}S_{\rm EMCS}=\frac{1}{16 \pi G}\int d^4 x\sqrt{-g}
\Big[R-F_{\mu\nu}F^{\mu\nu}-\beta\Big(\phi^2R+
6\partial_\mu\phi\partial^\mu\phi\Big)\Big],
\label{EMCS}
\end{eqnarray}
where the last term corresponds to a conformally coupled scalar action with parameter $\beta$. This action has no Weyl symmetry (under Weyl rescaling: $g_{\mu\nu}\to \Omega^2 g_{\mu\nu},~
\phi\to \phi/\Omega,~F_{\mu\nu}\to F_{\mu\nu}$) because  the Ricci scalar $R$ is present. In the absence of Maxwell term,
it corresponds to the Einstein-conformally coupled scalar theory which provides the BBMB black hole solution.
From Eq.(\ref{EMCS}), Bekenstein has obtained  the charged BBMB black hole solution~\cite{Bekenstein:1974sf} and then, Astorino has  found  the constant scalar hairy black hole solution~\cite{Astorino:2013sfa}.
In this work,  we choose $\beta=\kappa/6=4\pi G/3$.

 Einstein equation is derived  from  Eq.(\ref{EMCS}) as
\begin{equation} \label{nequa1}
G_{\mu\nu}=2 T^{\rm M}_{\mu\nu}+T^{\rm \phi}_{\mu\nu}
\end{equation}
with the Einstein tensor  is given by $G_{\mu\nu}=R_{\mu\nu}-Rg_{\mu\nu}/2$.
Here, two energy-momentum tensors for Maxwell theory and  conformally coupled scalar theory  are  defined by
\begin{eqnarray} \label{equa2}
T^{\rm M}_{\mu\nu}&=&F_{\mu\rho}F_{\nu}~^\rho- \frac{F^2}{4}g_{\mu\nu},\label{trace} \\
T^{\rm \phi}_{\mu\nu}&=&\beta\Big[\phi^2G_{\mu\nu}+g_{\mu\nu}\nabla^2(\phi^2)-\nabla_\mu\nabla_\nu(\phi^2)+6\nabla_\mu\phi\nabla_\nu\phi-3(\nabla\phi)^2g_{\mu\nu}\Big].\nonumber
\end{eqnarray}
At this stage, we check the traceless condition of $T^{{\rm M},\mu}_\mu=0$ easily. The Maxwell equation takes the form
\begin{equation} \label{maxwell-eq}
\nabla^\mu F_{\mu\nu}=0.
\end{equation}
Lastly, the scalar
equation is given by
\begin{equation} \label{ascalar-eq}
\nabla^2\phi-\frac{1}{6}R\phi=0.
\end{equation}
Even though $T^{\phi,\mu}_{\mu}=\beta \phi(-R\phi+6\nabla^2\phi)$ is not zero apparently, making use of the scalar equation (\ref{ascalar-eq}) leads to a traceless case of $T^{\phi,\mu}_{\mu}=0$.
This implies that  $T^{\phi,\mu}_{\mu}$ vanishes on-shell configurations.
On the other hand, taking the trace of the Einstein equation (\ref{nequa1}) together with (\ref{ascalar-eq}) leads to a vanishing Ricci scalar as
\begin{equation} \label{ricciz}
R=0.
\end{equation}
Plugging this back into Eq.(\ref{ascalar-eq}) simplifies  it  as
\begin{equation} \label{sscalar-eq}
\nabla^2\phi=0.
\end{equation}

One  finds  the constant scalar hairy (charged) black hole solution  given by~\cite{Astorino:2013sfa}
\begin{eqnarray}
&&ds^2_{\rm cshbh}=\bar{g}_{\mu\nu}dx^\mu dx^\nu=-N(r)dt^2+\frac{dr^2}{N(r)}+r^2d\Omega^2_2, \nonumber \\
&&N(r)=1-\frac{2m}{r}+\frac{Q^2+S}{r^2}, \label{bbmb2} \\
&&\bar{\phi}_\pm=\pm \sqrt{\frac{1}{\beta}}\sqrt{\frac{S}{Q^2+S}},~~\bar{A}_t=\frac{Q}{r}-\frac{Q}{r_+}. \nonumber
\end{eqnarray}
This is  derived mainly from solving the background Einstein equation
\begin{equation}
\bar{R}_{\mu\nu}=\frac{2\bar{T}^{\rm M}_{\mu\nu}}{1-\beta \bar{\phi}^2}\equiv 2\bar{T}_{\mu\nu} \label{back-eq1}
\end{equation}
with the traceless energy-momentum tensor
\begin{equation}
 \bar{T}^{\mu}~_\nu=\frac{Q^2+S}{r^4}{\rm diag}[-1,-1,1,1].\label{back-eq2}
 \end{equation}
 Considering an anisotropic matter with $T^\mu~_\nu={\rm diag}[-\rho,p,p_T,p_T]$, we have corresponding energy density and pressures
 \begin{equation}
 \rho=p_T=\frac{Q^2+S}{r^4},\quad p=-\frac{Q^2+S}{r^4}. \label{edap}
 \end{equation}
Imposing $N(r)=0$, the locations of outer and inner horizons
are determined  by
\begin{equation}
r_{\pm}=m\pm \sqrt{m^2-Q^2-S}.
\end{equation}

This black hole seems to have   a primary scalar hair and its geometry is similar to a non-extremal RN black hole except that the position of the horizon is shifted  by the presence of the scalar charge $S$. However, we wish to mention that a constant scalar  $\bar{\phi}_+(>0)$ depends on both $S$ and $Q$, implying that it is not strictly a primary hair.
We point out from Eq.(\ref{back-eq2}) that the background energy-momentum tensor  $\bar{T}^{\mu}~_\nu$  is always  traceless, and it satisfies both dominant and strong energy conditions whenever $S\ge -Q^2$~\cite{Astorino:2013sfa}.
This implies that two energy conditions violate for $S< -Q^2$.
Also, we observe from Eq.(\ref{back-eq1}) that for $1-\beta \bar{\phi}_+^2 \not=0(Q\not=0)$, the presence of a constant conformally coupled scalar field rescales the Newton  constant $G$ as $\tilde{G}=G/(1-\beta \bar{\phi}_+^2)=G(Q^2+S)/Q^2$.
This effective Newton  constant  should be  positive. This is possible  whenever  $S>-Q^2$ that corresponds to the condition for respecting  both dominant and strong energy conditions.
For $S=-Q^2$, one finds an unwanted case of $\tilde{G}=0$.
\begin{figure*}[t!]
   \centering
   \includegraphics{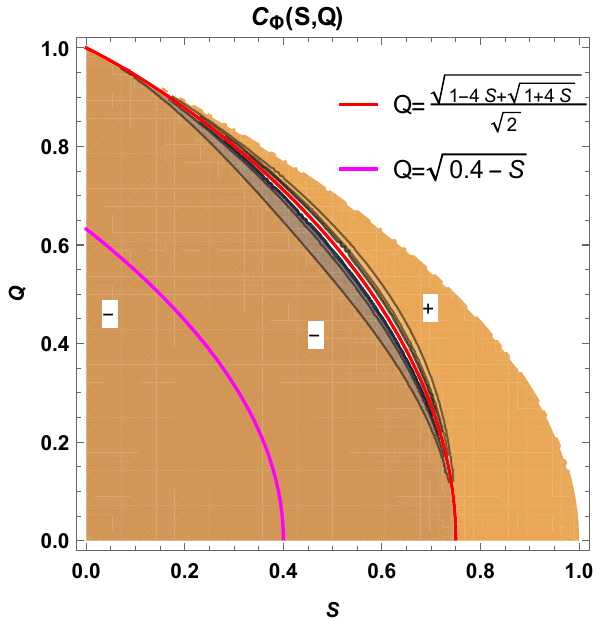}

\caption{Contour plot of heat capacity  $C_\Phi$ for SHRN black holes as function of scalar hair $S\in[0,1]$ and electric charge $Q\in[0,\sqrt{1-S}]$.
A red curve  represents a boundary (Davies curve: $C_\Phi:-\infty\to \infty$) between negative (unstable) region  and positive (stable) region.
A magenta curve $Q=\sqrt{0.4-S}$ denotes an upper limit for $0< S \lesssim 0.4$ [EHT($1\sigma$)] in SHRN black holes~\cite{Khodadi:2020jij}. }
\end{figure*}

Furthermore, its thermodynamic quantities  are  well established when replacing the Newton  constant $G$ with the effective Newton constant $\tilde{G}\not=0$:  ADM mass $M=\frac{m}{\tilde{G}}$  and entropy $S=\frac{A_+}{4\tilde{G}}$. We stress that two quantities are always positive for $\tilde{G}>0(S>-Q^2)$. These go to zero when the electric charge $Q$ approaches zero.  This suggests that the constant scalar hairy charged  black hole cannot radiate away its  charge $Q$ and
settle down to a constant scalar hairy (uncharged) black hole. The local thermodynamic stability can be determined by the heat capacity at constant electric potential
\begin{eqnarray}
C_\Phi&=&T\Big(\frac{\partial S}{\partial T}\Big)_\Phi=-\frac{2\pi r_+ Q^2(Q^2+2S)(r_+^2-Q^2-S)}{(Q^2+S)^2(r_+^2-Q^2-3S)},\label{h-eq1}
\end{eqnarray}
where the Hawking temperature is well-defined by
\begin{equation}
 T=\frac{r_+-r_-}{4\pi r_+^2}\ge0.
 \end{equation}
The zero temperature  is  allowed for an extremal black hole with $Q=\sqrt{1-S}$.
In this case, we confirm that the first-law of thermodynamics for black hole is satisfied.
$C_\Phi$ blows up at $Q=\sqrt{-S}$ and $\sqrt{1-4S+\sqrt{1+4S}}/\sqrt{2}$, while it vanishes at $Q=\sqrt{1-S}$ and $\sqrt{-2S}$. Its sign-dependence is shown in Fig. 1 for $0<S<1$ and Fig. 2 for $-5<S<0$.
A positive region (+: $C_\Phi>0$) represents a thermodynamically stable region, whereas  a negative  region ($-:~C_\Phi<0$) represents a thermodynamically unstable region. Two red curves denote Davies curves where $C_\Phi$ blows up. This information will be used for discussing the observationally  favored region based on the EHT($1\sigma$). 
\begin{figure*}[t!]
   \centering
   \includegraphics{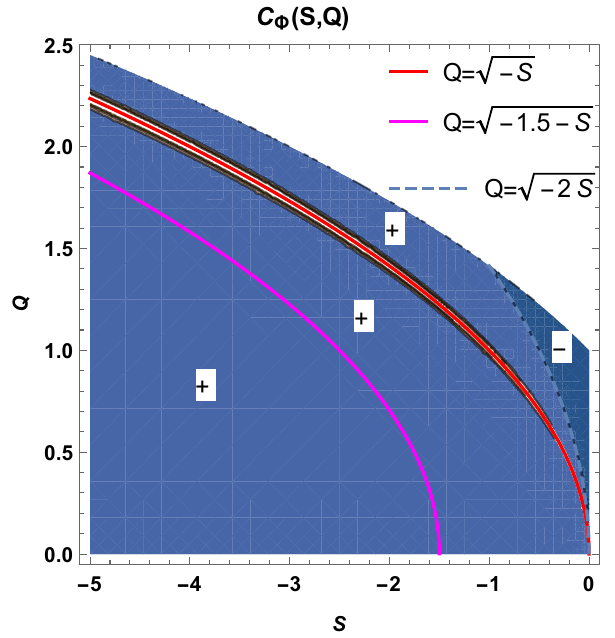}

\caption{Contour plot of heat capacity  $C_\Phi$ for scalar hairy charged and mutated-RN black holes as function of scalar hair $S\in[-5,0]$ and electric charge $Q\in[0,2.5]$.
A red curve $Q=\sqrt{-S}$ represents a boundary (Davies curve: $C_\Phi\to \infty$) between stable region (mutated-RN black holes) and stable region (scalar hairy charged black holes).
A magenta  curve $Q=\sqrt{-1.5-S}$ represents  a lower limit for $-1.5\lesssim S\lesssim 0$ [EHT($1\sigma)$] in mutated-RN black holes~\cite{Khodadi:2020jij}.
A dashed line $Q=\sqrt{-2S}$ represents a boundary ($C_\Phi=0$) between  stable  and unstable regions in scalar hairy charged black holes. }
\end{figure*}
\section{Photon ring bounds}

\begin{figure*}[t!]
   \centering
   \includegraphics{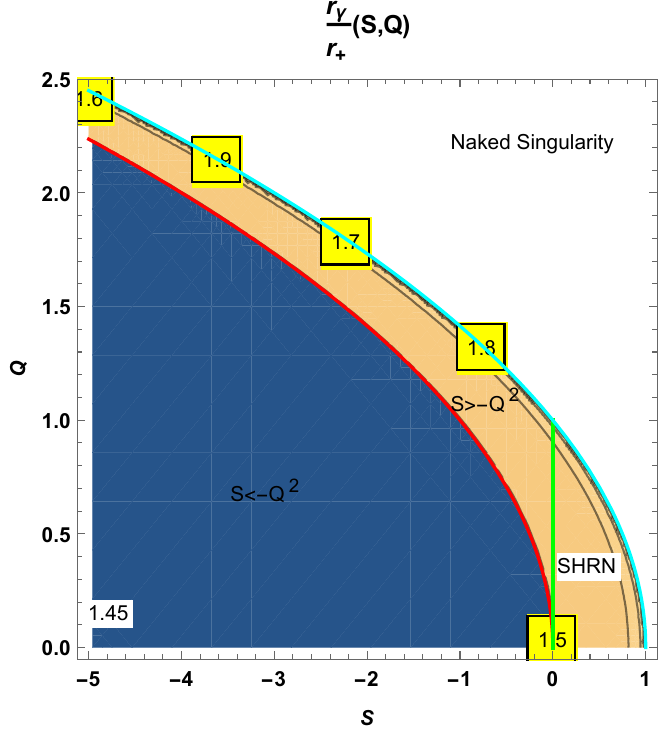}

\caption{Contour plot of  $\frac{r_\gamma}{r_+}$ as function of scalar hair $S\in[1,-5]$ and electric charge $Q\in[0,2.5]$.
A red curve $Q=\sqrt{-S}$ represents a boundary (Schwarzschild-like black hole with $\frac{r_\gamma}{r_+}=1.5$)  between scalar hairy charged black hole with  $S>-Q^2$ and  mutated-RN black hole  with $S<-Q^2$.
A cyan curve $Q=\sqrt{1-S}$ denotes a boundary (extremal  black hole with $\frac{r_\gamma}{r_+}=2$) between black holes with  $-Q^2<S<1$  and naked singularity with $S>1-Q^2$. The right of a green line represents the SHRN black holes with $0<S<1$ and $0<Q<\sqrt{1-S}$. At a point $(S=-5,Q=0)$, one finds  the lowest ratio of  $\frac{r_\gamma}{r_+}=1.45$.}
\end{figure*}

Before we proceed, we note that the lower bound on the photon ring radius $r_\gamma$  of black holes found  in Einstein gravity with a traceless energy-momentum tensor $T^\mu~_\nu={\rm diag}[-\rho,p,p_T,p_T]$ and $\rho(\ge0)\ge |p|,|p_T|$  is given by~\cite{Hod:2023jmx}
\begin{equation}
r_\gamma\ge 1.2 r_+. \label{l-b}
\end{equation}
Here, we wish to derive the ratio of $r_\gamma/r_+$  to obtain  the real bounds of  photon rings for  constant scalar hairy charged black holes
because the background energy-momentum tensor $\bar{T}^\mu~_\nu$ Eq.(\ref{back-eq2}) is  always traceless.
For this purpose, the effective potential for null geodesics is given by~\cite{daSilva:2023jxa}
\begin{equation}
V(r)=\frac{\sqrt{N(r)}}{r}.
\end{equation}
A photon ring  corresponds to a critical point of $V(r)$, that is,
\begin{equation}
V'(r)|_{r=r_\gamma}=0.
\end{equation}
In this case, one finds a photon ring radius with $m=1$ as
\begin{equation}
r_\gamma=\frac{3}{2}\Bigg[1+\sqrt{1-\frac{8}{9}(S+Q^2)}\Bigg].
\end{equation}
The ratio of $r_\gamma$ to $r_+$ is defined as
\begin{equation}
\frac{r_\gamma}{r_+}=\frac{\frac{3}{2}\Big[1+\sqrt{1-\frac{8}{9}(Q^2+S)}\Big]}{1+ \sqrt{1-(Q^2+S)}}.
\end{equation}
We find interesting three ratios from Fig. 3:  1.5 (Schwarzschild black hole) for $Q=\sqrt{-S}$, 2 (extremal black hole) for $Q=\sqrt{1-S}$, and  1.45 at a point ($S=-5,Q=0$).
Also, one obtains   $\sqrt{2}=1.414$ for $S=-\infty$, irrespective of $Q$.

We wish to classify all bounds according to the ratio of $r_\gamma/r_+$. \\
1) $1.2 \le r_\gamma/r_+< 1.414$ \\
We could not find any constant scalar hairy charged  black holes,  belonging to this bound. \\
2) $1.414\le r_\gamma/r_+< 1.5$ \\
 Mutated-RN black holes (Einstein-Rosen bridge: wormhole) with $S<-Q^2$ belong to this category.
 All of these black holes are not satisfied  with dominant and strong energy conditions, even though their scalar  is real. We would like to  mention that  all  mutated-RN  black holes do not possess positive ADM mass and entropy, but they are thermodynamically stable (see Fig. 2).
However, we point out that this bound was included  to set a  lower limit ($-1.5$) of  $-1.5\lesssim S\lesssim 0$ obtained from the angular-size of  M87*'s shadow  within EHT($1\sigma$) ($9.5 \lesssim d_{M87*} \lesssim 12.5$). This allowed region belongs to thermodynamically stable region~\cite{Khodadi:2020jij}.\\
3) $r_\gamma/r_+ =1.5(Q=\sqrt{-S})$ \\
 Schwarzschild-like black hole provides this ratio exactly  with an infinite scalar. This represents Davies curve where the heat capacity blows up. \\
 4) $1.5 < r_\gamma/r_+< 2.0(S>0)$ \\
All SHRN black holes with $0<S<1$ and $0<Q<\sqrt{1-S}$  satisfy  this bound. Their scalar  is always real. One region of $Q<\sqrt{1-4S+\sqrt{1+4S}}/\sqrt{2}$ is thermodynamically unstable, while the other  of $Q>\sqrt{1-4S+\sqrt{1+4S}}/\sqrt{2}$ is thermodynamically stable (see Fig. 1).
We note that these black holes respect both  dominant and strong energy conditions and have positive ADM mass and entropy. It is worth noting  that   SHRN  black holes  are stable against full perturbations~\cite{Zou:2019ays}.
Importantly,  this bound   includes the angular-size of  M87*'s shadow within $1\sigma$ ($9.5 \lesssim d_{M87*} \lesssim 12.5$)~\cite{Khodadi:2020jij}. In this case, the EHT~\cite{EventHorizonTelescope:2019dse} sets an upper limit (0.4) of $0<S\lesssim 0.4$ for $Q\ll1$ (thermodynamically unstable region).
 \\
5) $1.5 < r_\gamma/r_+< 2.0(S<0)$ \\
This bound holds  for  all  scalar hairy charged black holes with $S>-Q^2$.
These black holes respect both  dominant and strong energy conditions, and have positive ADM mass and entropy.
Here, thermodynamically stable region ($Q<\sqrt{-2S}$) is bigger than thermodynamically unstable region ($Q>\sqrt{-2S}$) (see Fig. 2).
We note that  the scalar  $\bar{\phi}_+$ is imaginary.
So, this bound was  discarded for testing the angular-size of  M87*'s shadow~\cite{Khodadi:2020jij} and gravitational lensing effects~\cite{QiQi:2023nex}. However, these black holes  seem to be better physically  than mutated-RN black holes. These black holes may be located in EHT($1\sigma$) when considering M87*. \\
6) $r_\gamma/r_+ =2.0(Q=\sqrt{1-S})$ \\
The extremal  black holes belong to this case. Their temperature and heat capacity vanish. \\
 7) $r_\gamma/r_+> 2.0$ \\
 No such constant scalar hairy black holes are found  within this lower bound.
Consequently, all constant scalar hairy charged black holes respect the lower bound (\ref{l-b})  for photon rings.
\begin{figure*}[t!]
   \centering
   \includegraphics{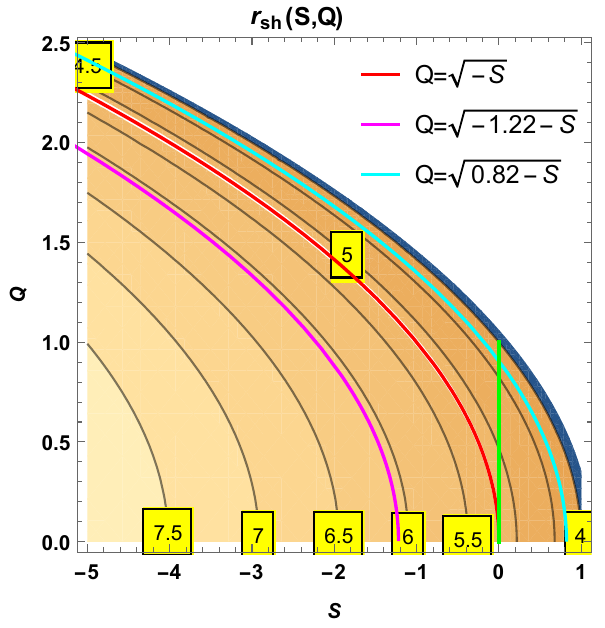}

\caption{EHT (shadow radius of M87*) constraints on $r_{\rm sh}$ as function of scalar hair $S\in[1,-5]$ and electric charge $Q\in[0,2.5]$.
The observationally favored region is given by $4.31\le r_{\rm sh} \le 6.08$~\cite{Xavier:2023exm} including Schwarzschild value $3\sqrt{3}\simeq 5.2$ (red curve).
Magenta  curve $Q=\sqrt{-1.22-S}$ represents  an upper  limit (6.08) in mutated-RN black hole, while  cyan curve $Q=\sqrt{0.82-S}$ denotes a lower limit (4.31) in SHRN black hole and   scalar hairy charged black hole with $S>-Q^2$.}
\end{figure*}

On the other hand, we introduce  the shadow radius defined by the critical impact factor as ~\cite{Vagnozzi:2022moj,daSilva:2023jxa}
\begin{equation}
r_{\rm sh}=\frac{r}{\sqrt{N(r)}}|_{r=r_\gamma}=\frac{\sqrt{2}\Big[3+\sqrt{9-8(Q^2+S)}\Big]}{\sqrt{4-\frac{3-\sqrt{9-8(Q^2+S)}}{Q^2+S}}}.\nonumber
\end{equation}
Its allowed region for the shadow radius is given by $4.31\le r_{\rm sh} \le 6.08$ with $m=1$~\cite{Xavier:2023exm}.
In this case, the observationally  favored region is shown in Fig. 4 which includes SHRN black hole, scalar hairy charged black hole with  $S> -Q^2$, and mutated-RN black hole.
This implies that  scalar hairy charged black hole with  $S> -Q^2$ will be located in EHT($1\sigma$) when considering the angular-size of  M87*'s shadow.

Finally, we mention  briefly the other  black hole solution obtained from the EMCS theory.
This is the charged BBMB black hole given by~\cite{Bekenstein:1974sf}
\begin{eqnarray}
&&ds^2_{\rm cBBMB}=-\Big(1-\frac{m}{ r}\Big)^2dt^2+\frac{dr^2}{\Big(1-\frac{m}{r}\Big)^2}+r^2d\Omega^2_2, \nonumber \\
&&\bar{\phi}_{B\pm}(r)=\pm \sqrt{\frac{1}{\beta}}\frac{S}{r-m},~~\bar{A}_t=\frac{Q}{r}-\frac{Q}{r_+},~~m=\sqrt{S^2+Q^2}, \label{bbmb1}
\end{eqnarray}
where $m(S,Q)$ is the mass of the black hole. The scalar  blows up at the horizon $r=m$ and it belongs to the secondary hair. The thermodynamical properties of the charged BBMB black hole are still  bad because the Hawking temperature  is always zero and the entropy becomes infinite. So, one argues that this might not  be  a physical black hole.
Furthermore, it was shown  that this black hole is unstable against the scalar perturbation~\cite{Bronnikov:1978mx,Onozawa:1995vu}.
Anyway, we fix  its ratio of  photon ring radius to horizon radius as $r_\gamma=2m$.
Also, its effective Newton constant at photon ring [$\tilde{G}|_{r=2m}=G/(1-\beta\bar{\phi}^2_{B\pm}(r))|_{r=2m}=GQ^2/(Q^2+S^2)$] takes a finite value, compared to $\tilde{G}|_{r=2m}=G/(1-\beta\bar{\phi}^2(r))|_{r=2m}\to \infty$ for the BBMB black hole~\cite{Shinohara:2021xry}.

\section{Discussions}

We have revisited  constant scalar hairy black holes  obtained from the EMCS   theory.
In this case, its background  energy-momentum tensor Eq.(\ref{back-eq2}) is always traceless.
These black holes are composed of SHRN black hole, scalar hairy charged black hole with  $S> -Q^2$, and mutated-RN black hole.
These black holes were candidates for testing whether they satisfy the lower bound of photon ring of $r_\gamma\ge 1.2 r_+$ proposed in~\cite{Hod:2023jmx}.
We have obtained  all real bounds of photon rings  for   these black holes and discussed physical and observational properties of real bounds.

The first and second ones  respect both dominant and strong energy conditions and they  show positive ADM mass and entropy. We remind the reader that  the scalar of the first is real, while the scalar of the  second is imaginary. The first (SHRN black hole) belongs to  a good category to test observational properties of M87* and SgrA* supermassive black holes. The observational region ($0<S\lesssim 0.4$) prefers thermodynamically unstable region  than  thermodynamically stable region.  The second (scalar hairy charged black hole) was discarded from the description for the angular-size of  M87*'s shadow~\cite{Khodadi:2020jij} and  gravitational lensing effects of M87* and SgrA* supermassive black holes~\cite{QiQi:2023nex} because its scalar is imaginary.   Here are two regions: thermodynamically stable ($Q<\sqrt{-2S}$) and  unstable regions ($Q>\sqrt{-2S}$).  We propose that the second should be included to test observational property of supermassive black holes because it belongs to a type of non-extremal RN black holes between Schwarzschild and extremal black holes~\cite{Vagnozzi:2022moj}.

On the other hand, the last (mutated-RN black hole) does not respect two energy conditions and it  show negative ADM mass and entropy.
So, this case has some handicaps to identify  physical black holes. We  note that the last is always  thermodynamically stable.
However, this  region was included as the description of EHT($1\sigma$) ($-1.5\lesssim S \lesssim 0$) for the angular-size of  M87*'s shadow~\cite{Khodadi:2020jij}.

Consequently, all constant scalar hairy charged black holes respect the lower bound (\ref{l-b})  for photon rings.
The observationally favored  region based on the shadow radius $r_{\rm sh}$  includes  SHRN black hole, scalar hairy charged black hole with  $S> -Q^2$, and mutated-RN black hole.
At this stage, it is not clear  which thermodynamical region (+ or $-$ heat capacity) is relevant to describing  observationally favored regions for supermassive black holes.

\end{document}